# A Gronwall Inequality Based Approach to Transient Stability Assessment for Power Grids

Qian Zhang, *Student Member, IEEE,*  Deqiang Gan, *Senior Member, IEEE*

*Abstract*—This paper proposes a novel Gronwall inequality-based method for transient stability assessment for power systems. The challenges of applying such methods to power systems are how to construct the differential inequality and how to treat its nonlinearity. By leveraging partial derivatives, a rotor angle difference inequality model is established, and the difficulty of nonlinearity of this model is solved by piecewise linearization. Based on our method, the upper bound of the rotor angle difference is given analytically, which can be used to estimate the stability boundary, i.e. the critical clearing time (CCT) of power systems. A case study on the IEEE 9-bus system shows the accuracy of the approach in early warning of transient instability for power grids.

## I. INTRODUCTION

In modern power systems, in order to ensure the safe and stable operation of the power network, it is necessary to estimate the transient stability margin, the result of the estimation is helpful for the early protection [1].

Many transient stability assessment tools have been proposed for power systems. Time-domain simulation is the most widely-adopted approach in the power industry, which can accurately identify the potential instabilities under given contingencies. The time-domain simulation has many advantages, such as visualization and flexibility of including the different grid components, but the computational burden and the inability to obtain the stability margin are the main limitations of this approach [2], [3]. An alternative method to transient stability analysis is energy function methods, which predict stability by comparing the system energy to a critical energy value [4]. While energy function methods can provide a quantitative measure of the stability region, the nonexistence of *general* energy functions for a lossy power grid introduces difficulty to the successful application of the method [5] [6], [7].

In recent years, some novel approaches were proposed in studying the second-order Kuramoto model [8]-[10], which was known closely related to the power network [11]. Different from conventional direct methods, the recently developed Kuramoto methods [12]-[15] do not rely on the complete information of the phase and frequency of all nodes, but focus on the diameters of the phase and frequency. The Gronwall inequality was first introduced to analyze the complete synchronization of Kuramoto oscillators with finite inertia in [13], where sufficient conditions were presented for asymptotic synchronization of oscillators with uniform coupling strength. This framework gives analytical results to the trapping region of the system, and the following paper [14] extended the model to oscillators with homogeneous frustration, which was similar to the lossy power system.

Compared to the second-order Kuramoto model shown in previous papers, the power network is more complex with heterogeneous coupling strength and frustration, which was fully considered in our work. To simplify this heterogeneous model, a partial derivative based approach is first introduced in this paper, which is more general than the trigonometric function-based simplify method in previous papers [13]-[15]. In addition, a novel piecewise linearization method is proposed to make the Gronwall inequality better suitable when applied to transient stability assessment of power systems. This approach can calculate the upper bound of rotor angle difference analytically, which is beneficial to the estimation of the CCT, it is also helpful in showing the relationship between system parameters and stability margin.

The rest of this paper is organized as follows: Section II constructs the rotor angle inequality model. Section III presents a Gronwall inequality based approach for transient stability assessment. Section IV validates the method using two test power systems, and Section V concludes the main results of this work.

## II. MODEL AND PRELIMINARIES

### A. Transient Stability Analysis Model of Power System

Consider a power network containing $n$ synchronous generators in the classical model, which is represented by the second-order swing equation:

$$\frac{M_i}{\omega_R}\ddot{\theta}_i + \frac{D_i}{\omega_R}\dot{\theta}_i = P_{Mi} - P_{Ei} \tag{1}$$

where $i \in \{1,2, …,n\}$, $M_i$, $D_i$, $\theta_i$, $P_{Mi}$, and $P_{Ei}$ are the inertia constant, damping constant, rotor angle, mechanical power, and electrical power of the $i$-th generator, respectively. $\omega_R$ is a reference angular frequency for the network.

For classical transient stability analysis, $P_{Mi}$ is supposed to be constant. In this paper, we assume a generator is modeled as a voltage source with constant voltage magnitude $E'_i$ connected to

Q. Zhang is with the Department of Electrical and Computer Engineering, Texas A&M University, College Station, USA. D. Gan is with the School of Electrical Engineering, Zhejiang University, Hangzhou, China. E-mail: zhangqianleo@tamu.edu, deqiang.gan@ieee.org

(*Corresponding author: D. Gan*)

the terminal node through transient reactance $x'_i$, and loads are modeled as constant impedances.

By applying the Kron reduction method [16], a power system was reduced to a network only containing $n$ generator nodes, where $P_{Ei}$ can be written as

$$P_{Ei} = |E'_i|^2 G_{ii} - \sum_{j=1, j \neq i}^{n} |E'_i E'_j Y_{ij}| \sin(\theta_j - \theta_i + \alpha_{ij}) \quad (2)$$

where $Y_{ij}$ represents the elements of complex admittance matrix $Y$ in the reduced network, i.e. $Y_{ij} = |Y_{ij}| \exp(j\varphi_{ij})$, and $G_{ii}$ is the real part of $Y_{ii}$. The parameter $\alpha_{ij} := \varphi_{ij} - \pi/2$ is determined by the line loss and active loads, which should not be ignored in a lossy power system.

**Definition 1** *(Bounded Synchronization)*: For a given $\zeta \in [0, \pi)$, system (1) achieves $\zeta$-bounded synchronization, if

$$\lim_{t \to \infty} \max |\theta_i(t) - \theta_j(t)| \leq \zeta, \quad i, j \in 1, 2, \ldots, n$$

The *Bounded Synchronization* is similar to the conception *Phase-Locked State* in paper [16].

**Definition 2** *(Frequency Synchronization)*: system (1) achieves frequency synchronization, if

$$\lim_{t \to \infty} |\dot{\theta}_i(t) - \dot{\theta}_j(t)| = 0, \quad i, j \in 1, 2, \ldots, n$$

The transient stability of power system is equal to bounded synchronization in terms of angle stability. Therefore, in practice, the maximum rotor angle difference less than 180° is taken as the criterion to evaluate the transient stability of the system, which is consistent with industry practice and has been found by utility engineers to be acceptable [23].

*B. Rotor Angle Difference Model*

Supposing the power system has homogeneous damping, i.e. $D_i/M_i = \lambda$, which was widely used in [12]-[15], we can rewrite the model (1) and (2) in the following form:

$$\ddot{\theta}_i + \lambda \dot{\theta}_i = \Omega_i + \sum_{j=1, j \neq i}^{n} a_{ij} \sin(\theta_j - \theta_i + \alpha_{ij}) \quad (3)$$

where $\Omega_i = \omega_R (P_{Mi} - |E'_i|^2 G_{ii})/M_i$, $a_{ij} = \omega_R |E'_i E'_j Y_{ij}|/M_i$.

As for transient stability analysis, the maximum rotor angle difference of generators directly reflects the stability margin of the system. To present the rotor angle difference model, we introduce the following notation.

**Notation 1**:

$$\theta_m = \min_{1 \leq i \leq n} \theta_i, \quad \theta_M = \max_{1 \leq i \leq n} \theta_i, \quad D(\theta) = \theta_M - \theta_m, \quad \theta_{ij} = \theta_i - \theta_j$$

Then the maximum rotor angle difference satisfies:

$$\ddot{D}(\theta) + \lambda \dot{D}(\theta) = D(\Omega) + \sum_{j \neq M}^{n} a_{Mj} \sin(\theta_j - \theta_M + \alpha_{Mj}) \\ - \sum_{j \neq m}^{n} a_{mj} \sin(\theta_j - \theta_m + \alpha_{mj}) \quad (4)$$

where $D(\Omega) = \Omega_M - \Omega_m$.

Notice that the differential equation (4) has two state variables: $D(\theta)$ and $\dot{D}(\theta)$, which are not explicit on the right-hand side(R.H.S) of the equation. Some papers [13]-[15] used trigonometric function to approximate the R.H.S of (4) to a formula that only contains variable $D(\theta)$, but the model they analyzed merely has homogenous frustration (phase shift) $\alpha$ or even without frustration term.

In this paper, we try to simplify the R.H.S of (4) with a more general method, which allows the model to have any value of phase shift $\alpha_{ij}$. Firstly, we translate the R.H.S of (4) to the form below:

$$\text{R.H.S.} = D(\Omega) + a_{Mm} \sin(-D(\theta) + \alpha_{Mm}) \\ - a_{mM} \sin(D(\theta) + \alpha_{mM}) + \sum_{j \neq M, m}^{n} f_j(\theta_j) \quad (5)$$

where $f_j(\theta_j) = a_{Mj} \sin(\theta_{jM} + \alpha_{Mj}) - a_{mj} \sin(\theta_{jm} + \alpha_{mj})$, which implies:

$$f_j(\theta_j) = a_{Mj} \sin(\theta_{jm} - D(\theta) + \alpha_{Mj}) - a_{mj} \sin(\theta_{jm} + \alpha_{mj}) \quad (6)$$

Noticing that for any given $D(\theta)$, the function $f_j(\theta_j)$ is determined by $\theta_{jm}$, which belongs to $[0, D(\theta)]$. Then we have:

$$\frac{\partial f_j(\theta_j)}{\partial \theta_{jm}} = a_{Mj} \cos(\theta_{jm} - D(\theta) + \alpha_{Mj}) - a_{mj} \cos(\theta_{jm} + \alpha_{mj}) \quad (7)$$

Based on the discussion of the sign of the derivative, we can easily calculate the maximum value of $f_j(\theta_j)$, which consists of the trig function of $D(\theta)$. Firstly, the partial derivative of $f_j(\theta_j)$ can be modified below:

$$\frac{\partial f_j(\theta_j)}{\partial \theta_{jm}} = (a_{Mj} \sin(D(\theta) - \alpha_{Mj}) + a_{mj} \sin \alpha_{mj}) \sin \theta_{jm} \\ + (a_{Mj} \cos(D(\theta) - \alpha_{Mj}) - a_{mj} \cos \alpha_{mj}) \cos \theta_{jm} \quad (8)$$

where $0 \leq \theta_{jm} \leq D(\theta) < \pi$.

Introduce auxiliary angle:

$$\varphi_0 = \arctan \frac{a_{Mj} \cos(D(\theta) - \alpha_{Mj}) - a_{mj} \cos \alpha_{mj}}{a_{Mj} \sin(D(\theta) - \alpha_{Mj}) + a_{mj} \sin \alpha_{mj}},$$

**Case A**: $\varphi_0 \geq 0$

The derivative of $f_j(\theta_j)$ has the following results:

$$\begin{cases} \partial f_j / \partial \theta_{jm} < 0, & 0 \leq \theta_{jm} < \pi - \varphi_0 \\ \partial f_j / \partial \theta_{jm} > 0, & \pi - \varphi_0 < \theta_{jm} < \pi \end{cases}$$

which means the maximum value of $f_j(\theta_j)$ can be represented as:

$$f_{j\max} = \max \{f_j(0), f_j(D(\theta))\} \\ = \max \{-a_{Mj}/m_M, -a_{mj}/m_m\} \cdot \sin(D(\theta)) \quad (9)$$

**Case B**: $\varphi_0 < 0$

The derivative of $f_j(\theta_j)$ is more complicated:

$$\begin{cases} \partial f_j / \partial \theta_{jm} > 0, & 0 \leq \theta_{jm} < -\varphi_0 \\ \partial f_j / \partial \theta_{jm} < 0, & -\varphi_0 < \theta_{jm} < \pi \end{cases}$$

If $D(\theta) < \varphi_0$, the maximum value of $f_j(\theta_j)$ is:

$$f_{j\max} = -a_{mj}/m_m \cdot \sin(D(\theta)) \quad (10)$$

otherwise, the maximum value is:

$$f_{j\max} = a_{Mj}\sin(\varphi_0 - D(\theta) + \alpha_{Mj}) - a_{mj}\sin(\varphi_0 + \alpha_{mj}) \quad (11)$$

After combining each trig term in (9)-(11) by trigonometric equations, we have:

$$\text{R.H.S.} \leq a_{Mm}\sin(-D(\theta) + \alpha_{Mm}) - a_{mM}\sin(D(\theta) + \alpha_{mM})$$
$$+ D(\Omega) + \sum_{j \neq M,m}^{n} f_{j\max} = \Gamma - L\sin(D(\theta) + \phi) \quad (12)$$

where $f_{j\max}$ can be obtained from (9)-(11), and $\Gamma$, $L$ and $\Phi$ are constants depending on system parameters with $\Gamma > 0, L > 0$.

**Remark**: Because of $\varphi_0$ in (11) is determined by $D(\theta)$, which means the parameters $\Gamma$, $L$ and $\Phi$ in equation (12) may have slightly different values in different intervals in some special case, but this change only affect the continuity of (12) in one point, and the monotonicity and concavity of the function are not affected, which means the idea of piecewise linearization in Section III still work.

## III. MAIN RESULTS

The power system model (3) shared almost the same mathematical expression with the second-order Kuramoto oscillators model [8], and some new results in these two research areas were inspired by each other [11]-[12]. Our main results are also trying to bridge the gap between these two different studies.

### A. Gronwall Inequality

**Lemma 1** (*Gronwall Inequality* [18]) Let $\psi : [0,T] \to \mathbb{R}$ be a nonnegative differentiable function, and $C(t)$ is a nonnegative integrable function. For every $t \in [0,T]$, if

$$\psi'(t) \leq C(t)\psi(t) \quad (13)$$

Then, we have

$$\psi(t) \leq \psi(0)\exp(\int_0^t C(\tau)d\tau) \quad (14)$$

The swing equations in transient stability analysis are second-order differential equations, which do not satisfy the condition of classical Gronwall Inequality in *Lemma1*.

*Theorem 1* to be described shortly extends Gronwall Inequality to second-order form, which was partly proposed in [13]. Before *Theorem 1* is presented, we first introduce the following Lemma:

**Lemma 2** (*Comparison Theorem* [18]) Let the function $U(t)$, $V(t)$ be differentiable in $J_0 : \xi < t < \xi + e$ and let the following hold:

(a) $U(t) < V(t)$ for $\xi < t < \xi + \varepsilon$ ($\varepsilon > 0$);

(b) $U'(t) < V'(t)$ in $J_0$.

Then:

$$U(t) < V(t) \text{ in } J_0.$$

**Theorem 1** (*Second-order Gronwall Inequality*) Consider a nonnegative function satisfying the following second-order differential inequality:

$$a\ddot{y} + b\dot{y} + cy + d \leq 0$$
$$y(0) = y_0, \dot{y}(0) = y_1 \quad (15)$$

where $a$, $b$, $c$ and $d$ are constants with $a > 0$. Then we have

(1.1) if $b^2 - 4ac > 0$

$$y(t) \leq (y_0 + d/c)e^{-v_1 t} +$$
$$a\frac{e^{-v_2 t} - e^{-v_1 t}}{\sqrt{b^2 - 4ac}} \times (y_1 + v_1 y_0 + \frac{2d}{b - \sqrt{b^2 - 4ac}}) - \frac{d}{c} \quad (16)$$

where $v_1 = \frac{b + \sqrt{b^2 - 4ac}}{2a}$, $v_2 = \frac{b - \sqrt{b^2 - 4ac}}{2a}$.

(1.2) if $b^2 - 4ac \leq 0$ and $b \neq 0$

$$y(t) \leq e^{-\frac{b}{2a}t}\left[y_0 + \frac{4ad}{b^2} + \left(\frac{b}{2a}y_0 + y_1 + \frac{2d}{b}\right)t\right] - \frac{4ad}{b^2} \quad (17)$$

(1.3) if $b^2 - 4ac \leq 0$ and $b = 0$

$$y(t) \leq -\frac{d}{2a}t^2 + y_1 t + y_0 \quad (18)$$

**Proof**: We focus on condition (1.3) in this paper, while the proof of (1.1) and (1.2) can be referred to Kuramoto oscillators-related study in [13]. When $b = 0$, inequality (15) can be simplified to the form below:

$$\ddot{y} \leq -cy/a + d/a \quad (19)$$

Notice that $c > 0$, because of $b^2 - 4ac \leq 0$ and $a > 0$, which means we have:

$$\ddot{y} \leq d/a \quad (20)$$

Based on *Lemma 2*, we can directly integrate both sides of inequality (20), i.e.:

$$y(t) \leq -\frac{d}{2a}t^2 + y_1 t + y_0 \quad (21)$$

### B. Piecewise Linearization

Because of the existence of the *sin* term, inequality (12) has strong nonlinearity. Unlike traditional linearization methods [13]-[15], we propose a novel piecewise linearization approach in this paper. To illustrate this method concisely, we suppose $\Gamma$, $L$ and $\Phi$ change little for different $D(\theta)$. Let $\gamma = D(\theta) + \phi$, inequality (12) can be modified to:

$$\ddot{\gamma} + \lambda\dot{\gamma} \leq \Gamma - L\sin\gamma \quad (22)$$

Taking $\gamma \in [0, 2\pi]$ for example, the *sin* function within a period is divided into four segments, i.e. $[0, \pi/2)$, $[\pi/2, \pi)$, $[\pi, 3\pi/2)$ and $[3\pi/2, 2\pi]$. As exhibited in Fig 1, the original *sin* function is piecewise linearized by connecting the endpoints of the function on these four intervals.

**Case 1:** $\gamma \in [0, \pi/2)$

$$\ddot{\gamma} + \lambda\dot{\gamma} \leq -L\sin\gamma + \Gamma \leq -\frac{2L}{\pi}\gamma + \Gamma \quad (23)$$

which means the parameters in (23) satisfy: $a = 1$, $b = \lambda$, $c = 2L/\pi$, $d = -\Gamma$.

**Case 2:** $\gamma \in [\pi/2, \pi)$

$$\ddot{\gamma} + \lambda \dot{\gamma} \leq -L\sin\gamma + \Gamma \leq \frac{2L}{\pi}\gamma - 2L + \Gamma \quad (24)$$

which means the parameters in (24) satisfy: $a = 1$, $b = \lambda$, $c = -2L/\pi$, $d = 2L - \Gamma$ and $b^2 - 4ac > 0$.

The piecewise linearization in $[\pi, 3\pi/2]$ and $[3\pi/2, 2\pi]$ also have the same results.

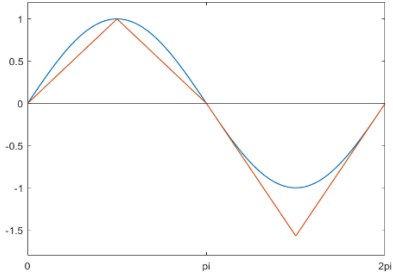

Fig. 1. Piecewise linearization diagram.

**Remark**: To better illustrate our core idea, we only presented a simple segmentation method in Fig.1. In general, the shorter segments we selected, the smaller the estimation error of the right hand side of (22), but the computation burden will also increase as more segments are picked.

### C. Switching Process

During the transient period of the power network after the occurrence of a fault, the maximum rotor angle difference will probably be across two segments in Fig 1, especially in strong inertia power systems. Because the parameters in differential inequality changed after switching to a different interval, the initial value of state variable in *Theorem* 1 needs to be redefined.

Because the system loses stability when $D(\theta)$ reaches 180°, and the parameter $\Phi$ is small in general system, the switching process for transient stability assessment mainly occurs between $[0, \pi/2]$ and $[\pi/2, \pi]$. Before giving the switching process calculation method, we first introduce some notations:

***Notation 2***:

$D(\theta_0)$ : Initial value of the maximum angle difference at fault clearing time;

$\dot{D}(\theta_0)$ : Initial value of the change rate of the maximum angle difference at fault clearing time;

$\tau$ : The time when the upper bound of maximum angle difference $\widehat{D}(\theta_\tau)$ calculated by Theorem 1 is reaching the turning point, i.e. $\pi/2$, while the real value of maximum angle difference is $D(\theta_\tau)$ and its change rate is $\dot{D}(\theta_\tau)$.

***Assumption***: The change rate of the maximum angle difference decreases from fault clearing time to time $\tau$, i.e. $\dot{D}(\theta_0) > \dot{D}(\theta_\tau)$.

***Remark***: This assumption is valid in power engineering, because the strategy of fault clearing is to increase the electrical power output of the generator, which will finally prevent the divergence of angle difference. The convexity of the first angle swing curve also supports this assumption.

From *Theorem 1* we have:

$$D(\theta) \leq (D(\theta_\tau) + d/c)e^{-v_1 t} + a\frac{e^{-v_2 t} - e^{-v_1 t}}{\sqrt{b^2 - 4ac}} \times (\dot{D}(\theta_\tau) + v_1 D(\theta_\tau) + \frac{2d}{b - \sqrt{b^2 - 4ac}}) - \frac{d}{c} \quad (25)$$

It is in general difficult to compute the state values, i.e. $D(\theta_\tau)$ and $\dot{D}(\theta_\tau)$, after clearing the fault for the transient stability assessment scenario. However, we can estimate the state values during switching process by using the initial value when the fault is cleared. The proof is as follows:

As $a > 0$, $v_1 > 0$, and $v_1 > v_2$, the right hand side of (25) is monotonically increasing with respect to $D(\theta_\tau)$ or $\dot{D}(\theta_\tau)$. Furthermore, because $\widehat{D}(\theta_\tau) > D(\theta_\tau)$ and $\dot{D}(\theta_0) > \dot{D}(\theta_\tau)$ from *Assumption*, we can replace $D(\theta_\tau)$ and $\dot{D}(\theta_\tau)$ with $\widehat{D}(\theta_\tau)$ and $\dot{D}(\theta_0)$ respectively, and the inequality still holds. Finally, we can replace (25) with the inequality below to calculate the maximum angle difference:

$$D(\theta) \leq (\widehat{D}(\theta_\tau) + d/c)e^{-v_1 t} + a\frac{e^{-v_2 t} - e^{-v_1 t}}{\sqrt{b^2 - 4ac}} \times (\dot{D}(\theta_0) + v_1 \widehat{D}(\theta_\tau) + \frac{2d}{b - \sqrt{b^2 - 4ac}}) - \frac{d}{c} \quad (26)$$

## IV. CASE STUDY

This section validates the proposed method using the IEEE 9-bus test power system. The initial conditions are obtained from solving power flow with MATPOWER, and the upper bound of the angle analytically estimated by Gronwall inequality based approach is compared to the numerical results, which is widely accepted in the power industry for transient stability analysis [1]. All the results are calculated in MATLAB.

The IEEE 9-bus system, detailed in [19], has three synchronous generators and three load buses with constant impedance type loads. Other type of loads, such as constant current load is also suitable in our method by Kron reduction [16]. The inertia constant parameter of generators 1, 2, and 3 is set to 47.28, 12.8, and 6.02, respectively. The damping parameter $\lambda$ in traditional synchronous generators is less than 3, averaging 0.2 in some test cases [19], [20]. For virtual synchronous generators(VSG), however, the damping parameter $\lambda$ is much bigger, typically more than 5, and can be changed adaptively in some papers [21], [22].

We consider a three-phase grounding fault occurred at bus 1 and the topology of the network will not change after fault clearing. Systems with different homogeneous damping, which means the system with parameter $\lambda = 8$ and $\lambda = 0.5$, is tested in this paper.

The proposed method can be used to estimate the upper bound of rotor angle difference under different fault durations. Fig 2(a) shows the numerical results with $\lambda = 8$ and the duration of fault is 0.6s, while there is no switching process during the first swing. Fig 2(b) shows the results with $\lambda = 0.5$ and the duration of fault is 0.2s, while the parameter during first swing in (25) was: $a=1$, $b=\lambda=0.5$, $c=45.5$, $d=-181.7$, which shifted to $c=-45.5$ and $d=108.7$ in (26) after switching at 0.32s. These two figures both show that our results are the upper bound of the numerical curves, which verifies the correctness of the method.

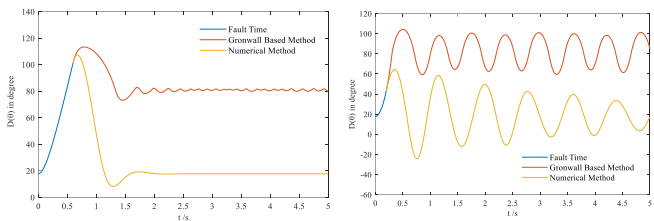

Fig. 2. Estimated upper bounds of rotor angle difference and the maximum rotor angle difference calculated by the numerical method in 9-bus system with parameter (a) $\lambda = 8$ (left hand), (b) $\lambda = 0.5$ (right hand).

If the rotor angle difference calculated by our method is less than 180°, then the system will maintain stable after fault clearing [23]. Based on this idea, we can also estimate the critical clearing time (CCT) of the fault, which is of vital importance for power system stability.

TABLE I. THE CCT BY DIFFERENT METHODS

| Parameter | CCT / s | |
|---|---|---|
| ($\lambda$) | Our Method | Numerical Result |
| $\lambda = 8$ | 0.66 | 0.81 |
| $\lambda = 0.5$ | 0.22 | 0.33 |

TABLE I shows the CCT calculated by numerical method and our method under different damping ($\lambda$). Since our approach focuses on the upper bound of rotor angle, the results are uniformly conservative, but this ensures the early alert before system breakdown. Based on our method, we can also derive indices to reflect the transient stability margin.

Taking $\mu = L / \Gamma$ as an index for example, TABLE II compares the influence of different line reactance on transient stability index $\mu$ and CCT in the 9-bus system with $\lambda=0.5$. The results verify the accuracy of $\mu$, which means the bigger $\mu$ is, the more stable system is.

TABLE II. THE INDEX AND CCT OF DIFFERENT LINE REACTANCE

| Line 5-7 reactance(p.u.) | Index($\mu$) | CCT(s) |
|---|---|---|
| 0.16 | 2.34 | 0.33 |
| 0.12 | 2.52 | 0.34 |
| 0.08 | 2.74 | 0.35 |

## V. CONCLUSION

A Gronwall inequality based approach to transient stability assessment of power system has been proposed in this paper. Compared with conventional energy function theory, our method directly reflect the stability region of power systems by analytically calculating upper bound of rotor angle difference. The effectiveness of the proposed method is tested in IEEE 9-bus system, where we quantitatively estimate the CCT to secure the transient stability of power system. A novel transient stability index is also derived and validated in this paper.